\documentclass[twocolumn,aps,superscriptaddress,prl
]{revtex4-1} 
\usepackage{amsmath,amssymb}
\usepackage{graphicx} 

\begin{document}
\title{Higher-order topology in honeycomb lattice with Y-Kekul\'e distortions}

\author{Yong-Cheng Jiang}
\affiliation{Research Center for Materials Nanoarchitectonics (MANA), National Institute for Materials Science (NIMS), Tsukuba 305-0044, Japan}
\affiliation{Graduate School of Science and Technology, University of Tsukuba, Tsukuba 305-8571, Japan}
\author{Toshikaze Kariyado}
\affiliation{Research Center for Materials Nanoarchitectonics (MANA), National Institute for Materials Science (NIMS), Tsukuba 305-0044, Japan}
\author{Xiao Hu}
\email{HU.Xiao@nims.go.jp}
\affiliation{Research Center for Materials Nanoarchitectonics (MANA), National Institute for Materials Science (NIMS), Tsukuba 305-0044, Japan}
\affiliation{Graduate School of Science and Technology, University of Tsukuba, Tsukuba 305-8571, Japan}

\date{\today}

\begin{abstract}
We investigate higher-order topological states in honeycomb lattice with Y-Kekul\'e distortions that preserve $C_{6v}$ crystalline symmetry.
The gapped states in expanded and shrunken distortions are adiabatically connected to isolated hexamers and Y-shaped tetramer states, respectively, where the former possesses nontrivial higher-order topology characterized by a $\mathbb{Z}_6$ invariant.
Topological corner states exist in a flake structure with expanded distortion where the hexamers are broken at the corners.
Our work reveals that honeycomb lattice with Y-Kekul\'e distortions serves as a promising platform to study higher-order topological states.
\end{abstract}

\maketitle


\textit{Introduction.}---Topological insulators~\cite{Hasan2010,Qi2011,Ando2013,Weng2015} are the materials that, while insulating in bulk, have conducting boundary states robust against nonmagnetic disorders due to the protection by time-reversal symmetry. 
Because of this property, they have attracted much attention for their potential applications in low-energy loss spintronic devices. 
The topological insulator in two dimensions (2D), also known as quantum spin Hall insulator (QSHI), was first theoretically proposed in graphene~\cite{Kane2005b} and then experimentally realized in HgTe quantum well~\cite{Bernevig2006,Konig2007}.
The topology of QSHI is captured by a $\mathbb{Z}_2$ invariant~\cite{Kane2005} protected by time-reversal symmetry, where spin-orbit coupling plays an essential role.
It is found that topological states can also arise from crystalline symmetries~\cite{Fu2011,Ando2015,Wu2015} even without spin-orbit coupling, where orbital degrees of freedom play a role similar to spin.
It has been proposed that deforming honeycomb lattice in a $C_{6v}$ symmetric way, namely with O-Kekul\'e distortions, can open an energy gap in the double Dirac dispersions with nontrivial topology induced by a band inversion between $p$- and $d$-like modes~\cite{Wu2015,Wu2016,Kariyado2017,Palmer2021}. In a simple tight-binding picture, the distortions are captured by hoppings with two different strengths, and the band inversion is induced by changing the ratio of the hopping amplitude.
This idea, which relies purely on crystalline symmetry rather than spin-orbit coupling, has been experimentally verified in spinless systems such as photonic crystals~\cite{Barik2018,Yang2018,Li2018,Parappurath2020,Shao2020,Wang2023}, acoustic systems~\cite{He2016}, and electronic artificial lattices~\cite{Gomes2012,Polini2013,Freeney2020}, through observing topological edge states.

With crystalline symmetries, the concept of topological insulator has been further extended to higher-order topology associated with boundary states of co-dimension larger than one~\cite{Benalcazar2017,Song2017,Schindler2018b,Benalcazar2019,Noh2018,Schindler2018a,Serra-Garcia2018,Peterson2018,Imhof2018,Kempkes2019,Xue2019,Mittal2019,Ni2019,ElHassan2019}, for example, 0D corner states in 2D materials, which arise from the mismatch between the number of electrons for charge neutrality and that preserving crystal symmetry~\cite{Song2017,Benalcazar2019}.
Corner states induced by nontrivial higher-order topology have been observed in a flake structure of honeycomb photonic crystal with O-Kekul\'e distortion~\cite{Noh2018}.
On the other hand, Y-Kekul\'e distortions with $C_{3v}$ symmetry have been experimentally observed in graphene recently, which was induced by the vacancies of copper substrate~\cite{Gutierrez2016}. 
This provides a novel platform for exploring electronic states associated with higher-order topology in real materials.

In this Letter, using a tight-binding model, we unveil that higher-order topological states can be achieved in honeycomb lattice by introducing Y-Kekul\'e distortions with the $C_{6v}$ crystalline symmetry preserved where half of the Y-shaped distortions point upward and the other half point downward. As in the case of the model with O-Kekul\'e distortions, Y-Kelul\'e distortions are effectively captured by introducing two types of hoppings. By tuning the distortion (or, the ratio of the hoppings), there appear gapped states in both expanded  and shrunken Y-Kekul\'e distortions, which are adiabatically connected to the set of isolated hexamers and the set of Y-shaped tetramers, respectively.
Especially, we reveal a nontrivial higher-order topology in the expanded distortion, which is characterized by a $\mathbb{Z}_6$ Berry phase with respect to the local twist of the Hamiltonian.
We show that the corresponding corner states can be observed in a flake structure where the hexamers are broken at the boundary.

\begin{figure}[t]
    \centering
    \includegraphics[width=0.48\textwidth]{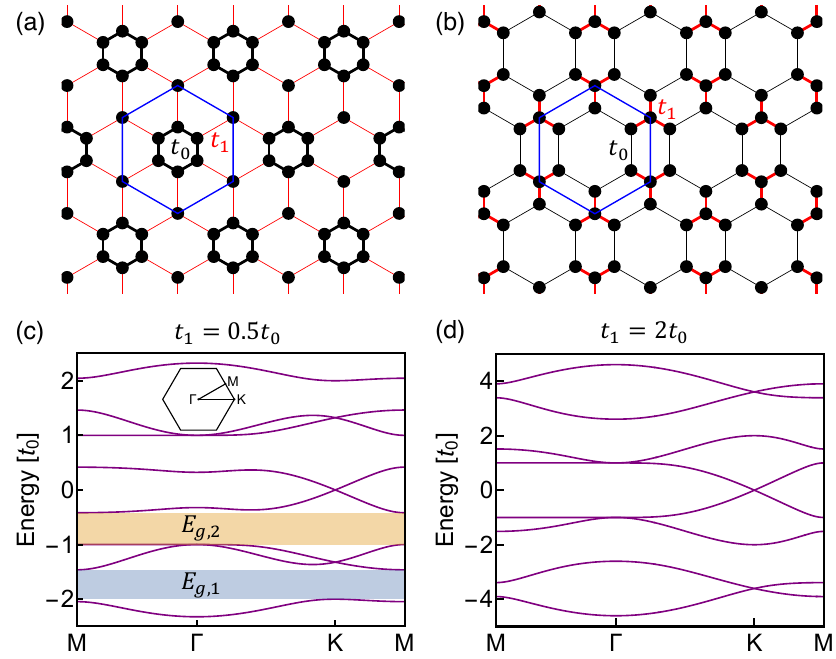}
    \caption{(a)~Schematic picture for honeycomb lattice with expanded Y-Kekul\'e distortion and global $C_{6v}$ symmetry. $t_0$ and $t_1$ are the nearest-neighbor hopping energies, and the highest-symmetric unit cell is shown by a blue hexagon. 
    The hexamer inside the unit cell with Hamiltonian~(\ref{eq:hamHex0}) is picked up for introducing local twist phases.
    (b)~Same as (a) except for shrunken distortion.
    (c)~Band structure of (a) with $t_1=0.5t_0$. The energy gaps highlighted in blue and orange are denoted as $E_{g,1}$ and $E_{g,2}$, respectively. Inset: Brillouin zone.
    (d)~Band structure of (b) with $t_1=2t_0$.}
    \label{fig:bulk}
\end{figure}

\textit{Model.}---We consider honeycomb lattice with expanded and shrunken Y-Kekul\'e distortions as shown in Figs.~\ref{fig:bulk}(a) and \ref{fig:bulk}(b), respectively. With these distortions, the lattice can be seen as a network of hexamers (black bonds) and tetramers (red bonds).
Both the expanded and shrunken structures respect $C_{6v}$ crystalline symmetry, with half of Y-shaped distortions pointing upward/downward.

In order to describe the electronic states in this system, we consider a tight-binding model for $p_z$ orbitals with nearest-neighbor (NN) hopping, which gives the Hamiltonian
\begin{equation}\label{eq:ham} 
    H = -t_0\sum_{\langle i,j\rangle}c^\dagger_ic_j -t_1\sum_{\langle i',j'\rangle}c^\dagger_{i'}c_{j'}, 
\end{equation}
where $t_0$ and $t_1$ are the NN hopping energies denoted by black and red bonds, respectively, in Figs.~\ref{fig:bulk}(a) and \ref{fig:bulk}(b). Because we focus on $p_z$-orbitals, this model is an eight-band model coming from eight sites in a unit cell.


We show the typical band structures for the expanded ($t_1=0.5t_0$) and shrunken ($t_1=2t_0$) Y-Kekul\'e distortions in Figs.~\ref{fig:bulk}(c) and \ref{fig:bulk}(d), respectively. 
Both band structures are symmetric with respective to $E=0$ due to the sublattice symmetry.
For the expanded distortion in Fig.~\ref{fig:bulk}(c), there are two energy gaps below the Fermi level ($E=0$), one between the first (counting from low energy) and the second bands and one between the third and the fourth bands, which we denote as $E_{g,1}$ and $E_{g,2}$, respectively.
For the shrunken distortion in Fig.~\ref{fig:bulk}(d), there is only one energy gap below the Fermi level, which is between the second and the third bands.

\begin{figure}[t]
    \centering
    \includegraphics[width=0.35\textwidth]{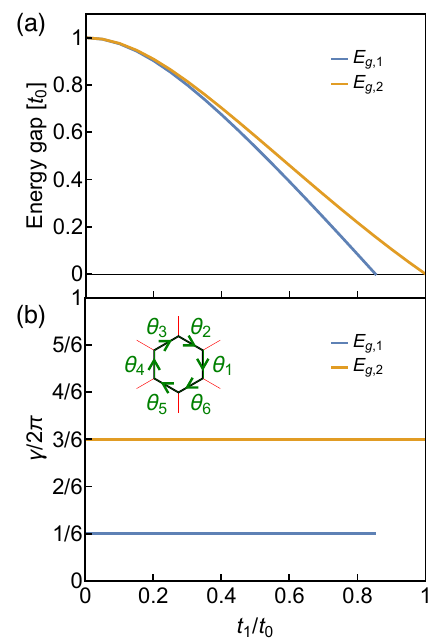}
    \caption{(a)~Energy gap as a function of $t_1/t_0$ for expanded Y-Kekul\'e distortion. 
    (b)~$\mathbb{Z}_6$ Berry phase $\gamma$ as a function of $t_1/t_0$. Inset: Twist phases for definition of Berry phase.}
    \label{fig:HOTI}
\end{figure}

In Fig.~\ref{fig:HOTI}(a) we show the band gaps $E_{g,1}$ and $E_{g,2}$ as functions of $t_1/t_0$ for the expanded distortion.
The energy gap $E_{g,1}$ remains open for $0\le t_1/t_0\lesssim0.85$.
For $0.85\lesssim t_1/t_0<1$, the system becomes metallic around $E=-2t_0$ with a negative indirect bandgap between the energy at the K point in the first band and the one at the M point in the second band.
The energy gap $E_{g,2}$ remains open for $0\le t_1/t_0<1$.
At $t_0=t_1$, both energy gaps close, since the system is nothing but the pristine honeycomb lattice without distortion.
For the shrunken distortion, a global energy gap is opened when $t_1/t_0 \gtrsim 1.63$.

\textit{Higher-order topology.}---We characterize the higher-order topology in our system using a Berry phase defined with respect to the local twist in Hamiltonian~\cite{Hatsugai2011,Kariyado2018a,Mizoguchi2019,Araki2020}. 
We first assume a sufficiently large supercell of our system, and pick up one of the hexamers (see Fig.~\ref{fig:bulk}(a)), for which the Hamiltonian is given by
\begin{equation}\label{eq:hamHex0}
    h_0 = -t_0\sum_{j=1}^{6} c^\dagger_{j+1}c_j + \text{H.c}.,
\end{equation}
given by NN-hopping terms within the hexamer.
The Hamiltonian for the rest part of the system is $H-h_0$.
Now we introduce local twist phases in the hexamer as shown in the inset of Fig.~\ref{fig:HOTI}(b), rendering Hamiltonian~(\ref{eq:hamHex0}) as
\begin{equation}\label{eq:hamHexTheta}
    h_0(\mathbf{\Theta}) = -t_0\sum_{j=1}^{6}  e^{i\theta_{j+1}}c^\dagger_{j+1}c_j + \text{H.c}.,
\end{equation}
with $\mathbf{\Theta}=(\theta_1,\theta_2,\theta_3,\theta_4,\theta_5)$ and $\theta_6=-\sum_{i=1}^5\theta_i$. 
Hamiltonian~(\ref{eq:ham}) for the whole system is then rewritten as
\begin{equation}
    H(\mathbf{\Theta}) = h_0(\mathbf{\Theta}) + (H-h_0).
\end{equation}
Denoting the ground state by $|\Psi(\mathbf{\Theta})\rangle$, the Berry phase is defined as
\begin{equation}\label{eq:gamma}
    \gamma = \int_{L_j} d\mathbf{\Theta} \mathbf{A}(\mathbf{\Theta}) \pmod{2\pi},
\end{equation}
where $\mathbf{A}(\mathbf{\Theta})= -i \langle \Psi(\mathbf{\Theta}) | \partial_{\mathbf{\Theta}} | \Psi(\mathbf{\Theta}) \rangle$ is the Berry connection.
Here $L_j=\mathbf{E}_{j-1}\rightarrow\mathbf{G}\rightarrow\mathbf{E}_j$ is a path in the $\mathbf{\Theta}$ parameter space with $\mathbf{E}_j=2\pi \hat{\mathbf{e}}_j$ and $\mathbf{G}=\frac{1}{5}\sum_{j=1}^5\mathbf{E}_j$, where $\hat{\mathbf{e}}_j$ $(j=1,\cdots,5)$ are the unit vectors and $\mathbf{E}_0=\mathbf{E}_6=\mathbf{0}$.
Due to the $C_6$ symmetry with respect to the center of the chosen hexamer which cycles $\theta_1$ to $\theta_6$ as shown in the inset of Fig.~\ref{fig:HOTI}(b) (equivalently in the parameter space the $C_6$ symmetry with respect to the central point $\mathbf{G}$ cycles $\mathbf{E}_1$ to $\mathbf{E}_6$), the summation of $\gamma$ in paths $L_1$ to $L_6$ is 0 modulo $2\pi$. Therefore, the $C_6$ symmetry guarantees a quantization of
\begin{equation}
    \gamma = 2\pi\frac{n}{6}, \quad n=1,\cdots,5 \pmod{2\pi},
\end{equation}
which indicates that the Berry phase is a $\mathbb{Z}_6$ index.

In Fig.~\ref{fig:HOTI}(b) we show the Berry phases as functions of $t_1/t_0$ for the expanded distortion calculated with a supercell including totally $7\times 7$ unit cells where periodic boundary conditions are applied.
For energy gap $E_{g,1}$ ($E_{g,2}$), the Berry phase remains quantized to $\pi/3$ ($\pi$) as long as the global energy gap does not close. 
This is intimately related to the fact that the ground state is adiabatically connected to the isolated hexamer cluster with $t_1=0$ at 1/6 (1/2) filling where the Wannier center is localized at the center of hexamer~\cite{Mizoguchi2019,Marzari2012}.

For the shrunken distortion, the gapped state between the second and third bands is adiabatically connected to the isolated Y-shaped tetramer cluster with $t_0=0$ at 1/4 filling. 
The state is topologically trivial since the Wanneir center for the shrunken distortion is located at the central lattice site of the tetramer, in contrast to the expanded distortion that the Wannier center does not coincide with any lattice sites.

\begin{figure}[t]
    \centering
    \includegraphics[width=0.48\textwidth]{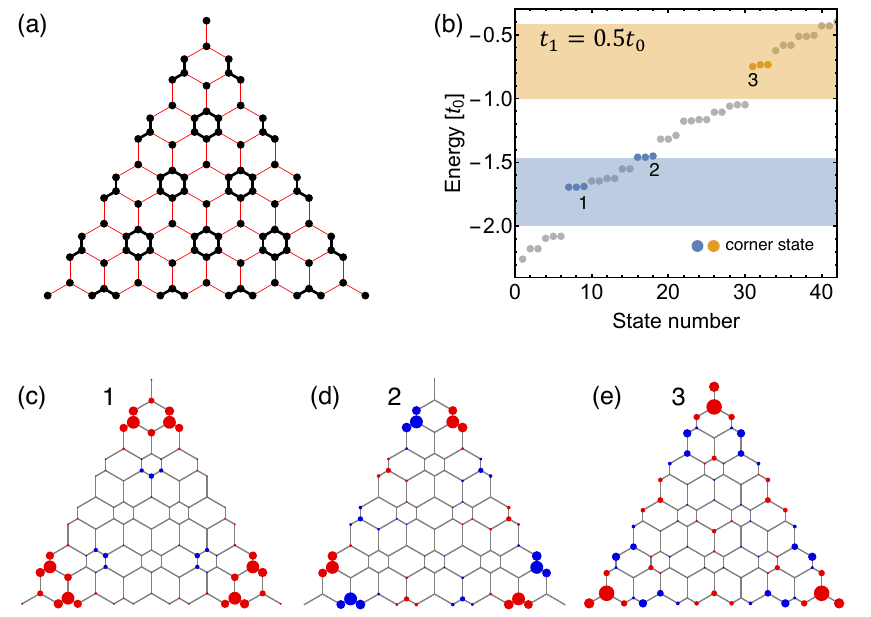}
    \caption{(a)~Flake structure with expanded Y-Kekul\'e distortion. 
    (b)~Energy spectrum of (a) with $t_1=0.5t_0$. The corner states in energy gaps $E_{g,1}$ and $E_{g,2}$ are highlighted in blue and orange dots, respectively, among which states 1 to 3 are labeled explicitly. 
    The energy gaps $E_{g,1}$ and $E_{g,2}$ for bulk are highlighted by blue and orange regions, respectively.
    (c)--(d)~Wave functions of corner states 1 to 3.}
    \label{fig:corner}
\end{figure}

\textit{Topological corner states.}---Associated with the higher-order topological index, in-gap states are expected to appear when the hexamers are broken at the boundary~\cite{Mizoguchi2019}. 
Here we consider a flake structure where the hexagons are broken at the boundary, as shown in Fig.~\ref{fig:corner}(a).
The energy spectrum with $t_1=0.5t_0$ is shown in Fig.~\ref{fig:corner}(b).
There appear two triplets of in-gap states in energy gap $E_{g,1}$ and one triplet of in-gap states in energy gap $E_{g,2}$. Due to the $C_{3v}$ symmetry of the flake, three corners are equivalent and give rise to the three degenerate states (with small degeneracy lifting due to finite-size effect). 
In each triplet we pick up the state with $+1$ $C_3$ eigenvalue, labeled as 1 to 3 in Fig.~\ref{fig:corner}(b), and plot their wave functions in Figs.~\ref{fig:corner}(c)--\ref{fig:corner}(e), which show a good localization at the corners.
Corner states 1 and 2 mainly localize at the two boomerang-like trimers, which are fragments of hexamer, with $t_0$ hopping.
The two trimers sit on the two sides of the mirror plane at the corner, and state 1 exhibits even mirror symmetry whereas state 2 shows odd mirror symmetry.
In contrast, corner state 3 mainly localizes at the Y-shaped tetramer with $t_1$ hopping, and thus carries a higher energy.
Since the tetramer sits at the mirror plane, only states with even mirror symmetry are allowed.

Since the Wannier centers in our system locate at the centers of hexamers and the boundary of nanoflake cuts the hexamers, there is a mismatch between the number of electrons for charge neutrality and that preserving crystal symmetry. This mismatch generates corner states, even when they are shifted energetically into bulk bands. However, detecting fractional charges would be very challenging if the corner states fall into bulk bands.


\textit{Discussion.}---In the flake structure with the expanded distortion shown in Fig.~\ref{fig:corner}(a), there also appear edge states between corner states 1 and 2, as shown in Fig.~\ref{fig:corner}(b).
These edge states originate from \textit{weak} first-order topology~\cite{Fu2007a}, indicated by the imbalance of parity index~\cite{Noh2018,Kariyado2018} between the $\Gamma$ point and two of the three distinct M points for the rhombic unit cell matching the corner of the flake.
The imbalance of parity index is one, resulting in one nearly dispersionless edge state. 

The energy difference between corner states and edge states in Fig.~\ref{fig:corner}(b) are larger than $0.04t_0$, corresponding to 0.1~eV in graphene ($t_0=2.7$~eV), which is large enough for scanning tunneling microscope to distinguish the corner states and the edge states. Moreover, it is possible to diminish the edge states within energy gap $E_{g,1}$ by making the central two hexamers complete out of four hexamers along each edge of the nanoflake in Fig.~\ref{fig:corner}(a), which leaves the Wannier centers associated with edge states unbroken along the edges~\cite{Kariyado2017}.

\textit{Conclusions.}---We study a honeycomb structure with Y-Kekul\'e distortions that maintains the $C_{6v}$ symmetry, and uncover its higher-order topology using a tight-binding model with nearest-neighbor hopping. 
For the expanded distortion where hopping energy $t_0$ within hexamers is stronger than hopping energy $t_1$ within Y-shaped tetramers, two energy gaps are found below the Fermi level.
Both energy gaps exhibit nontrivial higher-order topology as characterized by finite Berry phases, which is defined with respect to a local twist in one hexamer.
The Berry phase is quantized to a multiple of $2\pi/6$ due to the $C_6$ symmetry, and remains unchanged as long as the energy gap does not close.
The corresponding corner states can be observed in a flake where the hexamers are broken at the boundary.

Our model can be realized not only in real materials~\cite{Gutierrez2016}, but also in electronic artificial lattices, photonic crystals~\cite{Begum2023} and acoustic systems.
Even when next-nearest-neighbor hopping that breaks sublattice symmetry is taken into account, the topology remains unaffected provided that the energy gap does not close.
As elucidated in Ref.~\cite{Araki2020}, the nontrivial higher-order topology in our model characterized by the finite Berry phase in Eq.~(\ref{eq:gamma}) survives even in the presence of many-body interactions as long as the energy gap remains open. 
Our work shows that honeycomb lattice with Y-Kekul\'e distortions and $C_{6v}$ symmetry provides a promising platform to study the higher-order topological states.


\textbf{Acknowledgment} \quad This work is partially supported by CREST, JST (Core Research for Evolutionary Science and Technology, Japan Science and Technology Agency) (Grant Number JPMJCR18T4).

\bibliography{YKekule.bib}

\end{document}